\documentclass[aps,prl,twocolumn,article,superscriptaddress,showpacs]{revtex4-1}
\usepackage{graphicx}
\usepackage{dcolumn}
\usepackage{bm}

\begin{document}

\title{Atomic structure, energetics, and dynamics of topological solitons in indium chains on Si(111) surfaces}

\author{Hui Zhang}
\affiliation{Hefei National Laboratory for Physical Sciences at Microscale and Department of Physics, University of Science and Technology of China, 96 JinZhai Road, Hefei, Anhui 230026, China
}
\author{Jin-Ho Choi}
\affiliation{Department of physics, Hanyang University, 17 Haengdang-Dong, Seongdong-Ku, Seoul 133-791, Korea
}
\author{Yang Xu}
\affiliation{Hefei National Laboratory for Physical Sciences at Microscale and Department of Physics, University of Science and Technology of China, 96 JinZhai Road, Hefei, Anhui 230026, China
}
\author{Xiuxia Wang}
\affiliation{Hefei National Laboratory for Physical Sciences at Microscale and Department of Physics, University of Science and Technology of China, 96 JinZhai Road, Hefei, Anhui 230026, China
}
\author{Xiaofang Zhai}
\affiliation{Hefei National Laboratory for Physical Sciences at Microscale and Department of Physics, University of Science and Technology of China, 96 JinZhai Road, Hefei, Anhui 230026, China
}
\author{Bing Wang}
\affiliation{Hefei National Laboratory for Physical Sciences at Microscale and Department of Physics, University of Science and Technology of China, 96 JinZhai Road, Hefei, Anhui 230026, China
}
\author{Changgan Zeng}
\thanks{cgzeng@ustc.edu.cn}
\affiliation{Hefei National Laboratory for Physical Sciences at Microscale and Department of Physics, University of Science and Technology of China, 96 JinZhai Road, Hefei, Anhui 230026, China
}
\affiliation{ICQD, University of Science and Technology of China, Hefei, Anhui, 230026, China
}
\author{Jun-Hyung Cho}
\thanks{chojh@hanyang.ac.kr}
\affiliation{Department of physics, Hanyang University, 17 Haengdang-Dong, Seongdong-Ku, Seoul 133-791, Korea
}
\author{Zhenyu Zhang}
\affiliation{Materials Science and Technology Division, Oak Ridge National Laboratory, Oak Ridge, Tennessee 37831, USA
}
\affiliation{Department of Physics and Astronomy, University of Tennessee, Knoxville, Tennessee 37996, USA
}
\affiliation{ICQD, University of Science and Technology of China, Hefei, Anhui, 230026, China
}
\author{J. G. Hou}
\affiliation{Hefei National Laboratory for Physical Sciences at Microscale and Department of Physics, University of Science and Technology of China, 96 JinZhai Road, Hefei, Anhui 230026, China
}
\affiliation{ICQD, University of Science and Technology of China, Hefei, Anhui, 230026, China
}

\date{\today}

\begin{abstract}
Based on scanning tunneling microscopy and first-principles theoretical studies, we characterize the precise atomic structure of a topological soliton in In chains grown on Si(111) surfaces. Variable-temperature measurements of the soliton population allow us to determine the soliton formation energy to be $\sim$60 meV, smaller than one half of the band gap of $\sim$200 meV. Once created, these solitons have very low mobility, even though the activation energy is only about 20 meV; the sluggish nature is attributed to the exceptionally low attempt frequency for soliton migration. We further demonstrate local electric field enhanced soliton dynamics.
\end{abstract}


\maketitle


One-dimensional (1D) materials have attracted intensive attention due to their richness in physics and potential applications in nanoelectronics. The significantly enhanced interactions between charge, spin, and lattice in such 1D electronic systems could lead to exotic ground states, such as Luttinger liquid \cite{Kim1}, and charge density waves (CDWs) \cite{Yeom}. As prototype model systems, metals on semiconductor surfaces at low coverages could be self-organized to form atomic chains \cite{Barke}, which may exhibit 1D CDW states at low temperatures \cite{Snijders1}. The underlying physical properties of such collective electronic states in 1D systems, in turn, could be elucidated with atomic precision, for example by scanning tunneling microscopy (STM).

Besides the presence of exotic ground states, potentially even more intriguing are the elementary excitations of the 1D CDWs. Examples include the excitations of phase and amplitude (known as phason and amplitudon \cite{Gruner}), and the nonlinear topological excitation or soliton, which has been well studied in 1D conjugated polymers such as polyacetylene \cite{Su}. A soliton can be regarded as a local phase-flip boundary separating two energetically degenerate 1D CDW phases. The lattice displacements from the undistorted $\times$1 structure occur in opposite directions across a single soliton, thereby giving rise to a geometrical feature that is more complex than a single phase boundary. The formation energy of a soliton should also be less than one half of the band gap, otherwise electron-hole pairs would be more readily excited upon increasing the temperature. Solitons may possess spin-charge inversion properties, and act as the effective carriers that account for the high conductivity in conducting polymers \cite{Su}. However, earlier studies of solitons were based on ensemble average techniques \cite{Su}, and only very recently had some preliminary qualitative reports been made on the existence and dynamics of solitons using STM \cite{Morikawa,Lee1,Snijders2}. To date, a comprehensive quantitative study of topological solitary excitations at the atomic level remains a challenge.

In this Letter, we present a comprehensive study of the structure, energetics, and dynamics of topological solitary excitations at the atomic scale using a combination of STM and first-principles techniques. The precise atomic structure of the solitons in In chains on Si(111) surface is determined reliably. The formation energy and diffusion barrier of the solitons are estimated by examining the soliton population and thermal dynamics at different temperatures. We further demonstrate the capability of manipulating the soliton dynamics by applying an electric field between the STM tip and the solitons. Together, our findings represent the first insight into atomic-scale characteristics of topological excitations in such 1D systems.

The experiments were performed in an ultra-high vacuum system, equipped with an Omicron low temperature (LT) STM with a base pressure below 1$\times$10$^{-10}$ mbar. The Si(111) substrates (n-type, $\sim$0.001 $\Omega\cdot$cm) were cleaned using the standard ``flashing" recipe until the 7$\times$7 reconstruction was established at the surface. Then, the Si(111)-(4$\times$1)-In atomic chain structure was prepared by evaporating about one monolayer In onto a clean Si(111) surface at 700 K, followed by postannealing at the same temperature for 30 min.

As shown in Fig. 1(a), the STM image of the In chains on Si(111) at room temperature (RT) clearly shows the 4$\times$1 pattern. As the temperature is reduced, the system undergoes a reversible phase transition from a metallic 4$\times$1 phase to an insulating 8$\times$2 phase at a transition temperature around 130 K \cite{Tanikawa}. This is evidenced by the 8$\times$2 pattern in the STM image at 78 K (Fig. 1(b)). However, the structural formation of the 8$\times$2 phase is still controversial: One possible scheme is the pairing of the outermost In atoms, resulting in the formation of In trimers \cite{Kumpf}; the other includes additional shear distortion between two neighboring dimerized chains, leading to the formation of In hexagons \cite{Gonzalez1}. From previous density functional theory (DFT) calculations, it is found that the former structure is still metallic \cite{Cho}, while the latter is insulating \cite{Gonzalez1}. Here we acquired \emph{dI/dV} spectroscopy on the 8$\times$2 surface by adopting a locking-in technique, and the typical result is shown in Fig. 1(c). A fully-opened energy gap of about 0.2 eV is clearly observed, which supports the hexagon structural model.

\begin{figure}

\includegraphics{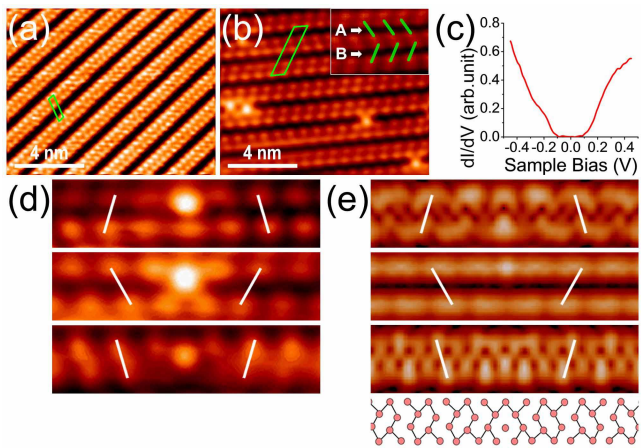}

\caption{(color online) Typical STM images of In chains on Si(111) surface at RT (a) and 78 K (b) at sample bias (\emph{V$_{s}$}) of -1.1 V and 1.5 V, respectively. The 4$\times$1 and 8$\times$2 unit cells are drawn in (a) and (b), respectively. The inset of (b) shows the two 4$\times$2 phases with alternating orientations composing the 8$\times$2 structure. (c) \emph{dI/dV} spectroscopes taken at the 8$\times$2 region. (d) Experimental STM images of the same soliton at \emph{V$_{s}$} of 0.6 V (top), 1.5 V (middle), and -0.6 V (bottom). (e) Simulated STM images of a soliton at \emph{V$_{s}$} of 0.6 V, 1.5 V, and -0.6 V from top to bottom, respectively. The adopted optimized soliton structure model is shown at the bottom.}
\end{figure}

The 8$\times$2 structure is composed of two 4$\times$2 chains with alternating orientations denoted as A and B phases in the inset of Fig. 1(b). The two 4$\times$2 phases are degenerate in energy with a gliding reflection symmetry. It is therefore possible that the A and B phases coexist in the same 4$\times$2 chain, and produce a phase boundary, namely, a topological solitary excitation between them.

In the LT STM images [see Fig. 1(b)], we find some bright features with well-defined uniform structure, which are absent at RT. More importantly, the orientations in the In chains are mirror-symmetric around the bright features, i.e., the bright features act as phase-flip boundaries separating the energy-degenerate A and B phases. Zoom-in STM images at different \emph{V$_{s}$} are shown in Fig. 1(d). The 4$\times$2 structure shows \emph{V$_{s}$}-dependent patterns \cite{Gonzalez2}, which are always mirror-symmetric around the phase boundaries. These phase boundaries are localized and preserve their structure even after diffusion, thus are likely to be originated from the formation of solitons.

In order to determine the precise soliton structure, we perform first-principles DFT calculations \cite{Hohenberg} within the generalized-gradient approximation \cite{Perdew}. All atoms are described by norm-conserving pseudopotentials \cite{Troullier}. On the basis of the 4$\times$2 hexagon model \cite{Gonzalez1}, we simulate the soilton structure by using a periodic slab geometry, where each slab contains six Si atomic layers (not including the Si surface chain) and the bottom Si layer is passivated by one H atom per Si atom. The vacuum spacing between these slabs is 10 \AA. The electronic wave functions are expanded in a plane-wave basis set with a cutoff energy of 15 Ry. The k-space integrations are done with four points in the surface Brillouin zone of the 4$\times$17 unit cell. Here we consider the phase boundary by employing the 4$\times$17 supercell where two solitons (separated by three In hexagons) are included because of the use of a periodic slab geometry. Based on our experimental evidence, we consider mirror symmetric In hexagons with respect to a plane of the phase boundary. The optimized soliton structure with a heart-shaped phase boundary is shown in Fig. 1(e).

For comparison with the STM observations, we simulate the constant-current STM images using the Tersoff-Hamann approximation \cite{Tersoff}, all at the charge density of $\rho$ = 10$^{-6}$ electrons/bohr$^3$. The results are displayed in Fig. 1(e). We find that for all the bias voltages considered, the orientations of the A and B phases as well as the separation of the bright spots agree very well between the theoretical simulations and experimental data. As shown in Fig. 1(e), the bright spots composing A and B orientations can be identified in terms of the two outer In atoms. It is interesting to note that, when the bias voltage changes from 0.6 V to 1.5 V, the orientations of A and B phases are swapped with each other in Fig. 1(d), which is caused by the slight lateral shifts of the bright spots between the 0.6 and 1.5 V images. The simulated images with \emph{V$_{s}$} = 0.6 and 1.5 V also reproduce well the features at the phase boundary as shown in the experimental ones. However for \emph{V$_{s}$} = -0.6 V, the simulated image at the phase boundary differs somewhat from the experimental image. This discrepancy is likely to be caused by the apparently different features in the simulated filled-state image, where weak protrusions originating from inner In atoms exist between adjacent paired bright spots, whereas the corresponding experimental image is unable to resolve such weak protrusions. Namely, the discrepancy is primarily due to the resolution limitation of the STM tip in resolving dark features \cite{Repp}.

\begin{figure}

\includegraphics{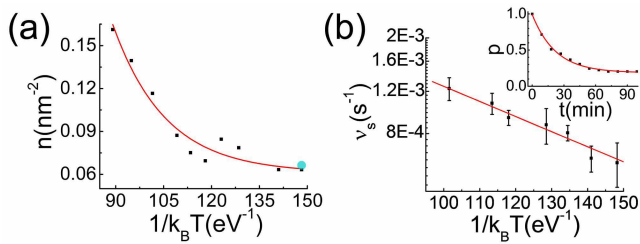}

\caption{(color online) (a) Temperature-dependent soliton density \emph{n} during warming-up. The large filled circle denotes \emph{n} at 78 K after cooling-down. (b) Temperature-dependent hopping rate $\nu_{s}$. The inset of (b) shows the time-dependent probability \emph{p} that solitons do not move during scanning at 86 K.}
\end{figure}

Temperature-dependent experiments are performed to explore the energetics of the solitons, with temperatures ranging from 78 K to 130 K. At temperature above $\sim$110 K, we observe the coexistence of the 4$\times$2 (8$\times$2) and 4$\times$1 phases \cite{Guo}, and focus on the soliton behavior in the 4$\times$2 (8$\times$2) phase. In determining the soliton formation energy, we measured the soliton density \emph{n} at various temperatures using the following procedure: At a given temperature, several well-defined 4$\times$2 (8$\times$2) regions were scanned, and the total number of isolated solitons was accumulated. To minimize the fluctuations in the soliton number, we ensured that the total soliton number counted at each temperature be about 500. The soliton density \emph{n} was then obtained by taking the total soliton number divided by the total area scanned. We checked the thermal reversibility by measuring the soliton population first by warming up and then by cooling down the systems. For the cool-down process, we only counted the soliton density at 78 K after cooling-down, and its value is close to that before warming-up as shown in Fig. 2(a). This cross check convincingly demonstrated that the soliton density does not depend on the thermal history, and the solitons (aside from that associated with \emph{n$_{0}$} discussed below) are indeed thermally excited.

Thus, the formation energy \emph{E$_{c}$} for a soliton can be evaluated by fitting the data using the formula \emph{n} = \emph{n$_{0}$}+\emph{n$_{1}$}exp(-\emph{E$_{c}$}/\emph{k$_{B}$}\emph{T}). We obtain \emph{$E_{c}$} of 61 $\pm$ 14 meV, less than one half of the band gap (200 meV). It means that the solitons, rather than the electron-hole pairs, are favored to be thermally excited \cite{Su}, consistent with the observations that solitons are stabilized in the In/Si(111) system. This further supports the soliton formation. Here the residual soliton density \emph{n$_{0}$} is estimated to be about 0.06 nm$^{-2}$. As defects are unavoidably present on the surface, some of the thermally excited solitons close to the defects could be stabilized to become immobile, and will not annihilate during cooling either, thereby leading to a non-zero \emph{n$_{0}$}. This observation has been further verified by going to 5 K in our STM observations: at this exceptionally low temperature, solitons can still be observed with a density close to \emph{n$_{0}$}. Using DFT calculations, we estimate the formation energy and the energy gap to be about 110 and 200 meV, respectively, close to the experimental values. In contrast to the experimental data, the calculated formation energy is slightly larger than half of the energy gap. This may be due to the well-known bandgap underestimation of the DFT calculations.

The importance of soliton migration lies in the aspect that solitons act as the effective carriers for high conductivity in conducting polymers. It is likewise intriguing to examine the soliton dynamics in the surface-based atomic chains scale as well, as such chains may serve as elemental building blocks in future nanoelectronics. From the time-dependent probability \emph{p}(\emph{t}) that solitons remain immobile during successive STM scanning, we can determine the hopping rate $\nu_{s}$ at a given temperature by fitting \emph{p}(\emph{t}) with \emph{p}(\emph{t}) = \emph{p$_{0}$}+(1-\emph{p$_{0}$})exp(-$\nu_{s}$\emph{t}). As an example, the fitting at 86 K is shown in the inset of Fig. 2(b). The non-zero \emph{p$_{0}$} is associated with \emph{n$_{0}$} discussed earlier, namely, some solitons close to pre-existing defects are trapped to be immobile. By fitting the temperature-dependent $\nu_{s}$ with the Arrhenius equation $\nu_{s}$ = \emph{A}exp(-\emph{$E_{a}$}/\emph{k$_{B}$}\emph{T}) \cite{Swartzentruber}, the diffusion barrier \emph{E$_{a}$} and the effective attempt frequency prefactor \emph{A} are estimated to be about 16 meV and 6$\times$10$^{-3}$ Hz, respectively.

Here we note that for typical surface diffusion, there exists an effective activation energy measuring the diffusion barrier, no matter what kind of diffusion mechanism involved. Usually the diffusion still follows the Arrhenius law, and the examples even include long jumps \cite{Senft} and defect-mediated diffusion \cite{Bussmann}. It is reasonable to fit the Arrhenius law to extract the activation energy in our case, and the excellent fitting displayed in Fig. 2(b) further confirms that an overall effective activation barrier against solition migration is well defined.

We note that anomalously low attempt frequency of the order of 10$^3$ Hz has been reported for Al atom diffusion on inhomogeneous Au(111) surface, which is associated with the low activation energy (30 meV) \cite{Fischer}. Generally observed trend is that the attempt frequency down shifts tremendously with shrinking diffusion barrier when the diffusion barrier is smaller than 100 meV \cite{Barth}, a phenomenon commonly known as the kinetic compensation effect \cite{Garn}. More importantly, unlike single atom or molecule, here the soliton is a collective topological excitation. Its diffusion must involve the rearrangements of many local bonds between the constituent atoms defining the soliton. Therefore, the diffusion of a soliton is in principle a collective behavior, and thus may possess an extremely low attempt frequency. We expect that the present quantitative study on the soliton mobility, with exceptionally low activation energy and attempt frequency, will stimulate future efforts on trying to uncover the precise atomistic mechanism for solition migration.

\begin{figure}

\includegraphics{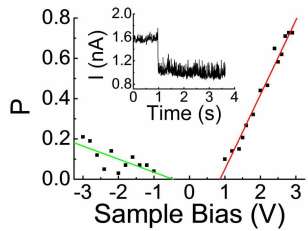}

\caption{(color online) The displacement probability per voltage pulse \emph{P} dependent on the pulse voltage. The inset is the I-t curve during a pulse with \emph{V$_{s}$} = 1.6 V.}
\end{figure}

We further explore the feasibility to enhance the soliton migration by applying a voltage pulse between the STM tip and the solitons at 78 K \cite{Ebert}. The experimental procedure is as follows: The STM tip is scanned over the surface with \emph{V$_{s}$} = 0.8 V and \emph{I} = 0.5 nA. Then it is moved to just over a soliton and a sample bias pulse with a period of 3.6 s is applied after disabling the feedback. The inset of Fig. 3 displays a recorded \emph{I}-\emph{t} curve when a pulse with \emph{V$_{s}$} = 1.6 V is applied. A steep decrease of the tunneling current is observed, indicating that the soliton moves away from its original position during the pulse.

Here we found that the displacement probability per pulse \emph{P} has a linear dependence on the bias voltage with about 80-100 events counted at each bias. Furthermore, there exists a clear polarity asymmetry: The displacement probability is much higher for positive bias than that for negative bias. For example, \emph{P} is almost 0.8 for a pulse with \emph{V$_{s}$} = 3.0 V, while only 0.2 for a pulse with \emph{V$_{s}$} = -3.0 V. This strong polarity asymmetry is possibly due to the enhanced Coulomb repulsion between the extra electrons and the n-type substrate when the electrons are injected into a soliton. It is noted that similar manifestation of the substrate doping effect on the electronic properties of surface layers has been previously reported \cite{Speer}.  These experimental findings clearly demonstrated the capability of enhancing the soliton dynamics at the atomic scale.

We may also acquire the effective mass of a soliton by adopting the standard procedure \cite{Su}. The soliton length 2\emph{l} (in unit of \emph{a}) is estimated to be roughly 5, from the full-width (along the chain direction) at half-maximum of the soliton feature in the STM image with \emph{V$_{s}$} = 1.5 V (after filtering the $\times$2 period modulation). The effective mass of the soliton can be obtained from \emph{M$_{s}$} = (4/3\emph{l})(\emph{u$_{0}$}/\emph{a})$^{2}$\emph{M}, where \emph{u$_{0}$} is the lattice distortion of the $\times$2 phase relative to the $\times$1 phase, and \emph{M} is the mass of the In$_{4}$ group. With \emph{u$_{0}$} = 0.35 \AA\ (estimated from the optimized structure) and \emph{l} = 2.5, \emph{M$_{s}$} is assessed to be about 3700 m$_{e}$ (m$_{e}$ denotes the electron rest mass), about twice of the mass of a hydrogen atom.

In polyacetylene, solitons have long length (2\emph{l} = 14), low diffusion barrier (2 meV), and light effective mass (6 m$_{e}$), acting as excellent carriers \cite{Su}. In contrast, the solitons in the In/Si(111) chains possess shorter length, higher effective mass, higher diffusion barrier. Together with the extremely low effective attempt frequency, they account for the sluggish nature of the solitons in the In/Si(111) chains.

In summary, we have characterized the precise atomic structure of a topological soliton in In chains grown on Si(111) surfaces, by combining STM measurements and theoretical simulations. The soliton formation energy is estimated to be $\sim$60 meV from the temperature-dependent soliton population, which is smaller than one half of the band gap. The solitons possess very low mobility, even though the activation energy is only about 16 meV; the sluggish nature is attributed to the exceptionally low attempt frequency for soliton migration. Local electric field enhanced soliton dynamics has been further demonstrated. Our work gains the first insight into characteristics of topological solitary excitations at the atomic scale in 1D electronic systems

This work was supported by NSFC (Grants No. 10974188, No. 50721091, and No. 10874164), ``One-hundred-person Project" of CAS, NKBRPC (Grant No. 2009CB929502), NCET, China Postdoctoral Science Foundation Funded Project (Grant No. 20100470837), National Research Foundation of Korea grant funded by the Korean Government (Grant No. KRF-2009-0073123), and in part by the Division of Materials Science and Engineering, Office of Basic Energy Sciences, Department of Energy, and NSF (Grant No. 0906025).

\end{document}